\documentclass{PoS}

\usepackage{mathtools}
\usepackage{amsfonts}
\usepackage{amssymb}
\usepackage{fdsymbol}
\usepackage{float}
\usepackage{graphicx}
\usepackage{multirow}
\usepackage{makecell}

\newcommand{\re}{\textmd{Re }}
\newcommand{\tr}{\textmd{tr }}

\title{The Anderson transition in QCD with $N_f=2+1+1$ twisted mass quarks: overlap analysis}
\ShortTitle{The Anderson transition in QCD with $N_f=2+1+1$ twisted mass quarks}

\author{\speaker{Lukas Holicki}
	\thanks{This work was supported by the Helmholtz International Center (HIC) for FAIR within the LOEWE initiative of the State of Hesse.}\\ 
	Institut f\"ur Theoretische Physik, Justus-Liebig-Universit\"at Gie\ss{}en, 35392 Gie\ss{}en, Germany\\
	\email{Lukas.Holicki@physik.uni-giessen.de}
}

\author{Ernst-Michael Ilgenfritz
		\thanks{Hospitality and support during a research visit at JLU within the HIC for FAIR visitor program are greatly acknowledged.} \\
        Bogolubov Laboratory for Theoretical Physics, Joint Institute for Nuclear Research, 141980 Dubna, Russia \\
        \email{ilgenfri@theor.jinr.ru}
}

\author{Lorenz von Smekal\\
	Institut f\"ur Theoretische Physik, Justus-Liebig-Universit\"at Gie\ss{}en, 35392 Gie\ss{}en, Germany\\
	\email{Lorenz.Smekal@theo.physik.uni-giessen.de}
}

\abstract{
	Chiral Random Matrix Theory has proven to describe the spectral 
properties of low temperature QCD very well. However, at temperatures 
above the chiral symmetry restoring transition it can not provide a global 
description. The level-spacing distribution in the lower part of the spectrum 
of the Dirac operator is Poisson-like. There the eigenmodes are localized in 
space-time and separated from the rest of the spectrum by a so-called mobility 
edge. In analogy to Anderson localization in condensed-matter systems with 
random disorder this has been called the QCD-Anderson transition. 
Here, we study the localization features of the low-lying eigenmodes of the 
massless overlap operator on configurations generated with $N_f=2+1+1$ twisted 
mass Wilson sea quarks and present results concerning the temperature 
dependence of the mobility edge and the mechanism of the quark-mode 
localization. We have used various methods to fix the spectral position of 
the delocalization transition and verify that the mobility edge extrapolates 
to zero at a temperature within the chiral transition region.
}

\FullConference{The 36th Annual International Symposium on Lattice Field Theory - LATTICE2018\\
		22-28 July, 2018\\
		Michigan State University, East Lansing, Michigan, USA.}

\begin{document}

\section{Introduction}
Since P. W. Anderson described the vanishing zero-temperature conductivity 
in imperfect crystals occurring when the quenched disorder exceeds some 
threshold~\cite{Anderson:1958vr}, transitions from extended to localized 
states have attracted much interest. Examples have been found in a variety 
of systems. In lattice gauge theory this has been extensively studied, 
beginning with SU(2) gauge theory
~\cite{PhysRevLett.74.3920,Kovacs:2009zj,Kovacs:2010wx,Bruckmann:2008xr}, 
as well as in QCD, both quenched and unquenched. The spectrum of the Dirac 
operator in the high-temperature phase consists of two coexisting regimes. 
It turned out that the issue of localization is closely connected to a
purely spectral feature like the loss of eigenvalue repulsion.

Below $T_c$ the Dirac operator spectrum is entirely described by Random 
Matrix Theory (RMT) \cite{Klein:2000pj,Verbaarschot:2000dy}, where the 
global symmetries of a theory are translated into universal spectral 
features and observables become averages over random matrices. 
These averages must be stripped of microscopic fluctuations, 
%%% stripped off ?
for example the level spacing distribution must be unfolded. 
Three-color QCD corresponds to the chiral Gaussian Unitary ensemble 
(GUE) with a Dyson index $\beta_D=2$~\cite{Halasz:1995vd}.

Above $T_c$, a lower end of the spectrum appears that can not be described 
with random matrices anymore. The eigenvalue spacing in this lower part is 
uncorrelated, and the corresponding eigenmodes are localized. 
The energy scale $\lambda_c$, that separates this region from the delocalized 
region is called ``mobility edge'', in analogy to the Anderson transition in 
condensed matter physics.

\begin{figure}[!ht]
\centering
\includegraphics[width=0.6\textwidth]{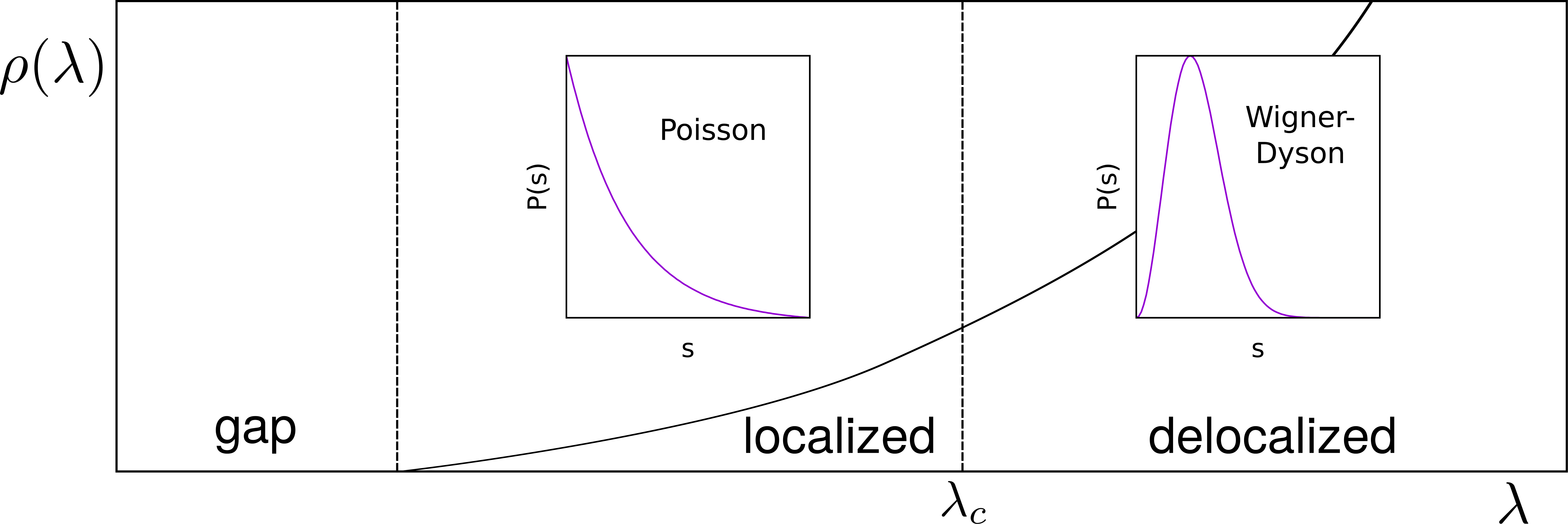}
\caption{The localized and delocalized spectral regimes of the QCD Dirac 
operator for $T>T_c$ and the corresponding level spacing distributions $P(s)$,
as predicted by Random Matrix Theory.}
%%% Does RMT also PREDICT the POISSON-like part ? 
\label{fig:sketch_mobility_edge}
\end{figure}

In quenched QCD the localized regime appears first at the deconfinement phase 
transition, whereas in full QCD it appears above the chiral phase transition.
In both cases, a so-called Banks-Casher gap opens in the spectral density $\rho(\lambda)$,
as schematically illustrated in Fig. \ref{fig:sketch_mobility_edge}.

In quenched QCD at high temperatures the spectral density of eigenstates 
(of the staggered Dirac operator) shows a square root behaviour 
~\cite{Damgaard:2000cx}, and the delocalization energy scale has been 
empirically identified using the inverse participation ratio 
(IPR)~\cite{Gavai:2008xe}. A clear correlation between the space-time 
density of the localized eigenmodes and the local Polyakov loop has been 
observed~\cite{Kovacs:2011tj}. The modes close to the mobility edge 
possess a multifractal structure \cite{Nishigaki:2013uya,Ujfalusi:2015nha}.
This has been shown for both full QCD as well as in models with orthogonal 
and symplectic universality classes \cite{Ujfalusi2015}. The critical exponent 
of this transition occurring in the Dirac operator spectrum has been found 
to agree with the critical exponent of the corresponding Anderson model 
~\cite{Pittler:2014qea,Giordano:2014pfa}. Fluctuations of the Polyakov loop 
and regions of exceptional self-duality have been identified to act as source 
for mode localization in the case of M\"obius Domain-Wall fermions 
~\cite{Cossu:2016scb}.  

On the background of SU(3) gauge field configurations, generated in full QCD 
with 2+1+1 dynamical fermions,
we have studied the spectrum and eigenmodes of the massless overlap operator,
~\cite{Neuberger:1997fp}
\begin{equation}
D = \frac{\hat{\rho}}{a}\left( 1 + \textmd{sgn} K \right) ,
\end{equation}
that exactly fulfills the Ginsparg-Wilson equation 
$\left\{ \gamma_5, D \right\} = \frac{a}{\hat{\rho}} D \gamma_5 D$,
where $\hat{\rho} \in (0,2]$ is an arbitrary scaling factor.
The sign kernel $\textmd{sgn} K = \frac{K}{\sqrt{K^\dag K}}$ was computed 
with a rational approximation while for the kernel $K$ we used the Wilson 
operator with a negative mass. In the continuum, the Dirac eigenvalues are 
purely imaginary, but on the lattice the eigenvalues are distributed on the 
Ginsparg-Wilson circle $ 2\textmd{Re }\lambda = \frac{a}{\hat{\rho}} | \lambda |^2 $. 
The continuum limit $a\to0$ is mimicked by stereographically projecting 
the eigenvalues onto the imaginary axis.

In order to set the scene for disorder and finite temperature, we have
used gauge field configurations generated by the tmfT collaboration 
(twisted mass at finite 
temperature)~\cite{Burger:2013hia,Burger:2015xda,Burger:2017xkz}
with $N_f=2+1+1$ flavors of dynamical Wilson fermions and using the 
Iwasaki gauge action. While strange and charm quark have physical masses, 
the pion mass is still unphysically large, $m_\pi \approx 370\textmd{ MeV}$, 
in the ensembles considered here.
%%% Irgendwann spaeter:
%%% To smooth short-range fluctuations within the gauge configurations, 
%%% these were cooled using Yang Mills Gradient Flow using the Iwasaki 
%%% action as well.

Our spectral computations were performed on two groups of ensembles with 
different lattice spacings, for several temperatures selected by $N_t$, 
see Table \ref{tab:lattice_setup}. The eigenmodes of the 
overlap operator on these gauge configurations were computed using the 
Implicitly Restarted Arnoldi method (IRAM).

\begin{table}[!ht]
\centering

\begin{tabular}{|c|c|c|c|}
	\hline
		%\multicolumn{4}{|c|}{ \makecell{\textbf{A370} \\ $N_s=24$ \\ $a = 0.0936 \textmd{ fm}$ \\ $L^3=11.336\textmd{ fm}^3$ \\ $m_\pi=364\textmd{ MeV}$ }} \\
		\multicolumn{4}{|c|}{ \makecell{\textbf{A370}: $N_s=24$, $a = 0.0936 \textmd{ fm}$ }} \\
	\hline
		$N_t$	& 	T / MeV 	&	number of conf. 	& 	$\frac{\textmd{modes}}{\textmd{conf}}$ \\
	\hline
		4		&	527.06	&	98	&	512		\\
		5		&	421.65	&	63	&	512		\\
		6		&	351.37	&	111	&	512		\\
		7		&	301.18	&	50	&	512		\\
		8		&	263.53	&	100	&	512		\\
		9		&	234.25 	&	101	&	512		\\
		10	&	210.82 	&	99	&	512		\\
	\hline
\end{tabular}
\hskip1cm
\begin{tabular}{|c|c|c|c|}
	\hline
		%\multicolumn{4}{|c|}{ \makecell{ \textbf{D370} \\ $N_s=32$ \\ $a = 0.0646 \textmd{ fm}$ \\ $L^3=8.834\textmd{ fm}^3$ \\ $m_\pi=369\textmd{ MeV}$ } } \\
		\multicolumn{4}{|c|}{ \makecell{ \textbf{D370}: $N_s=32$, $a = 0.0646 \textmd{ fm}$ } } \\
	\hline
		$N_t$	& 	T / MeV 	&	number of conf. 	& 	$\frac{\textmd{modes}}{\textmd{conf}}$ \\
	\hline
		3	&	1018.21	&	96		&	300		\\
			&			&			&			\\
		6	&	509.11	&	71		&	300		\\
			&			&			&			\\
		14	&	218.19	&	121		&	190		\\
	\hline
\end{tabular}

\caption{The tmfT lattice ensembles used in this work.}
\label{tab:lattice_setup}
\end{table}

\section{The moblity edge $\lambda_c$}

%%% definition of the mobility edge ?

Below the temperature of the chiral crossover, the Dirac spectrum of QCD is 
entirely dominated by its global symmetries and can be classified with a 
Dyson universality class. In this regime universal features of QCD are well 
described by Gaussian ensembles of random matrices.
%% of certain type. 
Above $T_c$, however, RMT fails to describe the lowest part of the spectrum, 
%%% Matrizen enthalten einen nichtzufälligen additiven Teil ?
%%% Kann RMT etwas über Form und Lokalisation aussagen ? 
that eventually contains localized eigenmodes and where the level spacings 
%%% levels oder level spacings ?
%%% level spacings sind unkorreliert, level repulsion ist nicht beobachtet.
are uncorrelated to each other.
%%% , e.g. that there is no level repulsion observed.
Recent studies in quenched QCD imply, that the localized part of the spectrum
completely disappears, when the deconfinement transition temperature is 
approached from above~\cite{Kovacs:2017uiz}.

\begin{figure}[!ht]
\centering
\includegraphics[width=0.45\textwidth]{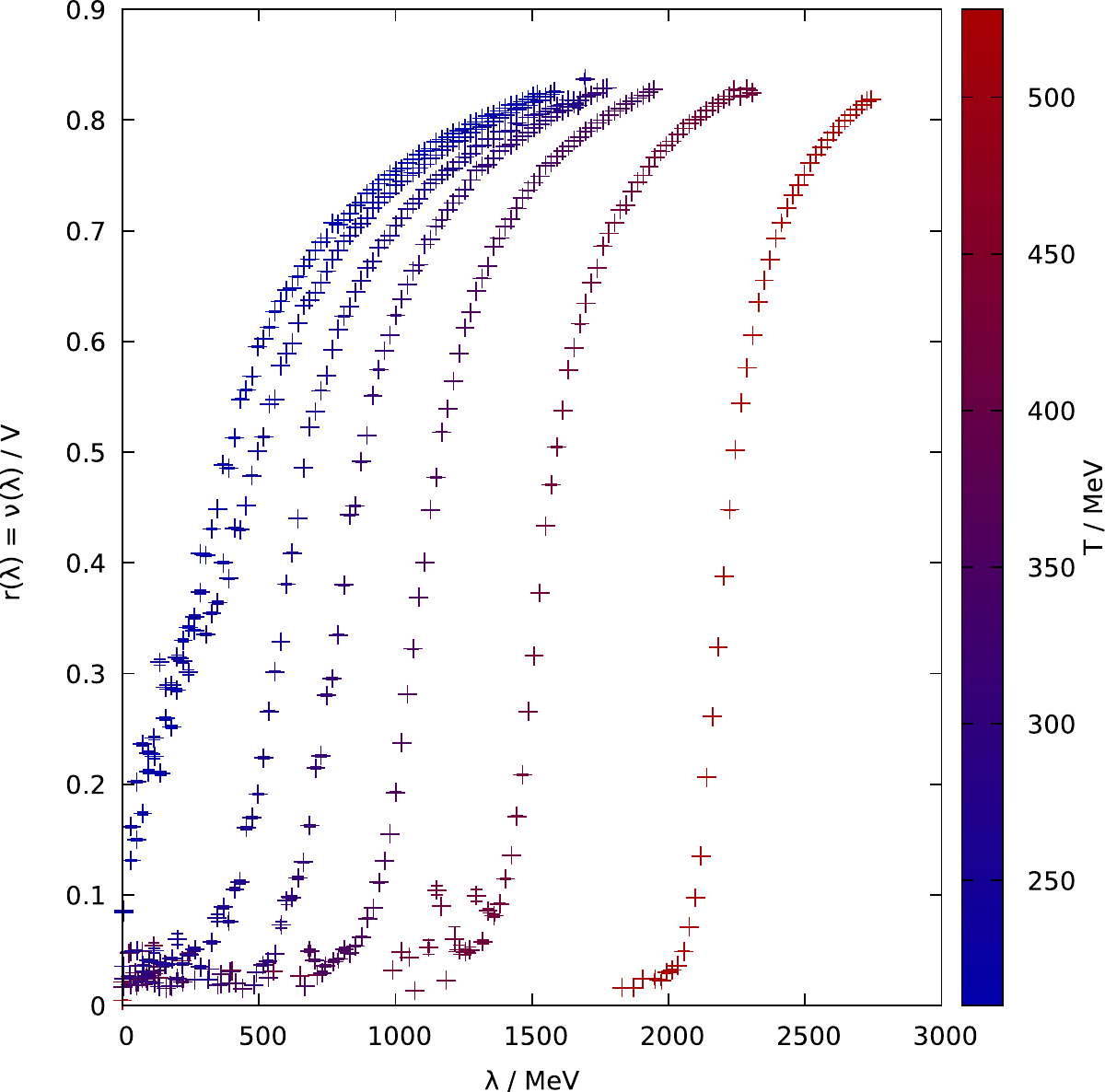}
\includegraphics[width=0.45\textwidth]{me_pr.pdf}
\caption{\textit{Left}: The relative eigenvector volume inside bins of width
$\Delta\lambda= 0.005 / a$ for several temperature ensembles of the A370 
configurations. \textit{Right}: The mobility edge $\lambda_c(T)$ is linear 
in the temperature. The temperature of the chiral transition was taken 
from \cite{Burger:2015xda}.}
\label{fig:rel_eigenmode_volume_and_lambdac_over_temperature}
\end{figure}

To get an appropiate estimate for the mobility edge $\lambda_c$, we have 
computed the inverse participation ratio 
(IPR) for each eigenmode (characterized by eigenvalue $\lambda$
and wave function $|\psi\rangle$)
\begin{equation}
\nu^{-1} = \int d^4x~\langle \psi(x) | \psi(x) \rangle^2 .
\end{equation}
If an eigenmode is uniformly extended over the whole volume, 
$\langle\psi(x)|\psi(x)\rangle=\frac{1}{V},~ \forall x$ the participation ratio
becomes $\nu=V$, while for a maximally localized mode at $x_0$ with a scalar
density $a^4 \langle\psi(x)|\psi(x)\rangle=\delta_{x,x_0}$ one gets $\nu=a^4$. 
We can then define a relative eigenmode volume 
$r(\lambda) = \frac{\nu(\lambda)}{V}$. On the left side of Fig.
\ref{fig:rel_eigenmode_volume_and_lambdac_over_temperature} we present 
$\langle r(\lambda) \rangle$, averaged over small bins with bin size 
$\Delta\lambda= 0.005 / a$.

From a fit to $\langle r(\lambda) \rangle$ we determined the inflection 
point of the curve and used this as an estimate for $\lambda_c$.
This coincides with the point where fluctuations within 
each bin are maximal. Other criteria applied to the relative eigenmode 
volume have yielded qualitatively the same results.

The estimates for $\lambda_c$ from the A370 and the D370 meta-ensembles 
imply, that the mobility edge vanishes at the temperature of the chiral 
crossover, that was determined in \cite{Burger:2015xda}. It also agrees 
well with a linear 
%%% fit as function of the temperature, which can be seen 
function of the temperature, which can be seen on the right side of Fig.
\ref{fig:rel_eigenmode_volume_and_lambdac_over_temperature}. In particular there is no evidence of curvature, as observed with staggered quarks in \cite{Giordano:2014pfa}.

\section{Localization and Polyakov loop}

Localization seems to be caused by some gauge-invariant objects
that apparently do not exist below $T_c$ but become 
%%% increasingly 
influential with increasing temperature. Such objects would represent 
the analog to the disorder in the condensed-matter Anderson models.
A good candidate for such a local quantity is defects in the Polyakov loop
\begin{equation}
l(\vec{x}) = \frac{1}{N_c} \tr {\cal P} e^{\int_0^{\beta} dt A_4(\vec{x},t)},
\end{equation}
the trace of the local holonomy (or its real part $\re l(\vec{x})$). 

In \cite{Kovacs:2011tj} and \cite{Bruckmann:2011cc} the connection between 
the local real-valued
$l(\vec{x})$ and the eigenmodes of the Dirac operator has been studied 
for SU(2) gauge theory. It has been found that the localized modes are 
trapped in sinks of $l(\vec{x})$. 
%%% $\re l(x)$ for SU(2) is real-valued
In \cite{Giordano:2015vla} an Anderson-like Ising model has been designed to 
mimick the effect of the Polyakov loop able to cause localization. 
In \cite{Cossu:2016scb} this correlation has been studied in full QCD with 
$N_f=2$ M\"obius Domain-Wall fermions.

To draw the connection of the modes (which are defined with respect to 
the original gauge field configurations) with the Polyakov loop of the latter
requires some amount of gauge field smearing.
To remove UV-fluctuations, the gauge configurations were smoothed with the
help of Gradient flow (with respect to the Iwasaki gauge action that was used here, until a 
flow time $a^2\tau=4.0$). 

We observe that the regions, where $\re l(x)$ is small, form clusters of
small size and are spatially localized at the same positions where the chiral 
zero modes in the corresponding configuration are localized.
%, see figure \ref{fig:ploop_clusters}.
%\begin{figure}[!ht]
%\centering
%\includegraphics[width=0.3\textwidth]{img/qcd/{ploop.cluster.D45.32.6.0048.T4.0}.png}
%\includegraphics[width=0.3\textwidth]{img/qcd/{D45.32.6.gf_T4.000e+00.0048.001-0}.png}
%\caption{$T=509 \textmd{ MeV} \approx 2.77 T_c$: Left: Spatial regions where 
%$\re l(x) < 0.3$ come in clusters, coinciding with the spatially projected 
%chiral zero modes existing in the given configuration. 
%Right: The pseudoscalar density $\langle \psi(x) | \gamma_5 | \psi(x) \rangle$ 
%of all zero modes in this configuration.}
%\label{fig:ploop_clusters}
%\end{figure}

\begin{figure}[!ht]
\centering
	\includegraphics[width=\textwidth]{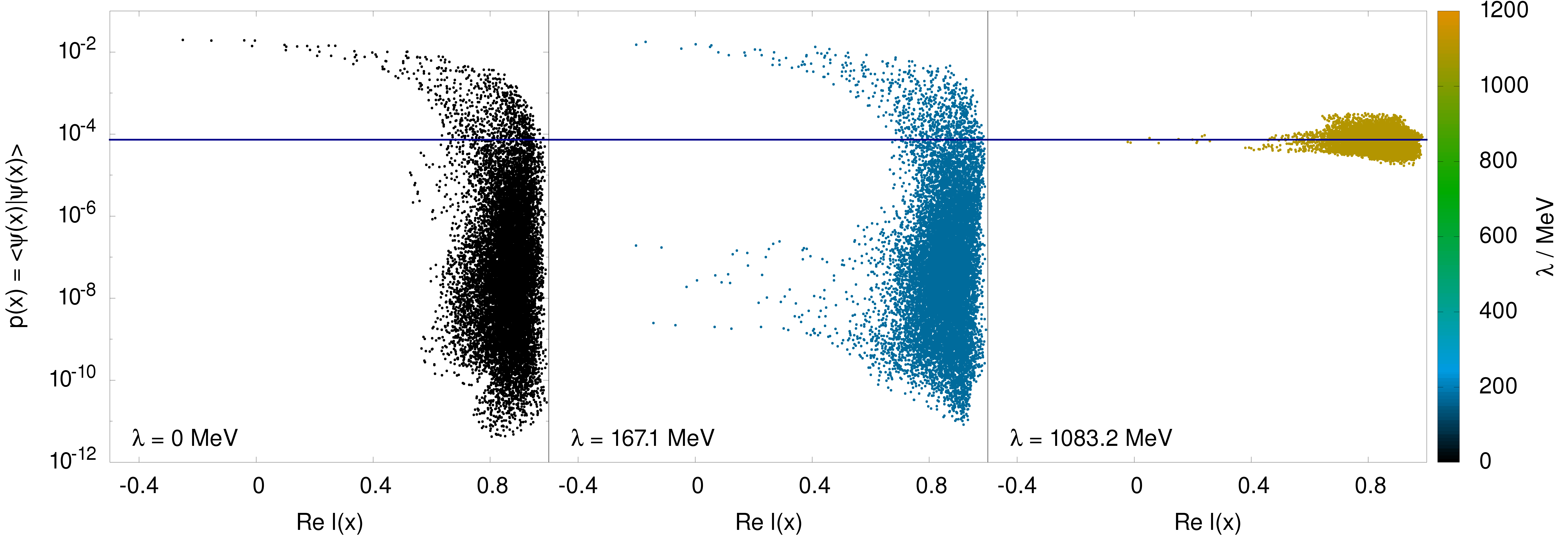}
\caption{Three typical eigenmodes correlated with the local Polyakov loop at 
$T=351 \textmd{ MeV} \approx 1.91 T_c$. 
\textit{Left}: A chiral zeromode ($\lambda=0\textmd{ MeV}$). 
\textit{Middle}: A localized nonzero mode ($\lambda=167.1\textmd{ MeV}\approx0.24\lambda_c$). 
\textit{Right}: An extended nonzero mode ($\lambda=1083.2\textmd{ MeV}\approx 1.53\lambda_c$).}
\label{fig:ploop_localization_A60.24.6}
\end{figure}

We have studied for three classes of modes the correlation between the scalar 
eigenmode density $p(x) = \langle \psi(x) | \psi(x) \rangle$ and the local 
Polyakov loop, which is shown in Fig. \ref{fig:ploop_localization_A60.24.6} 
for $T=351 \textmd{ MeV}$.
%%% and \ref{fig:ploop_localization_D45.32.6} for 
%%% $T=509 \textmd{ MeV}$. Here 
The horizontal line indicates the density $p(x)=\frac{1}{V}$. If a mode is 
homogenously extended throughout the whole volume, its density would collapse 
on this line. We observe, that both the zero modes and the localized non-zero 
modes are strongly anticorrelated with the real part of the Polyakov loop (left and middle panel of figure \ref{fig:ploop_localization_A60.24.6}). 
In regions, where $\re l(x)$ is large, these modes are strongly supressed. 
Delocalized modes, in contrast, appear to be mostly indifferent with respect
to the Polyakov loop, and are spread in a small stripe around the 
$\frac{1}{V}$-line (right panel).
Indeed, negative real part values of the Polyakov loop appear to be acting as 
source for localization, in analogy to the random on-site potential in the 
condensed matter Anderson model. 
%{\bf to be demonstrated by placing the center of localized modes into the 
%complex Polyakov loop plane at the appropriate place, see Cossu.}
%	Ich glaube Cossu plottet nicht das Zentrum (dazu waeren wahrscheinlich
%	die Fluktuationen zu irrefuehrend), sondern einen Mittelwert ueber viele
%	Moden.
%% This conjecture
%{\bf conjecture or finding ?}
%	Conjecture. Wir haben aus unseren Bildern geschlossen das der
%	Polyakov Loop eine bestimmte Rolle uebernimmt und eine
%	Analogie formuliert, das ist kein objektives Ergebnis
%	sondern der Vorschlag zu einer Interpretation.
This conjecture agrees very well with earlier findings \cite{Cossu:2016scb} based on Domain-Wall quarks.

%%% weglassen
%%% \begin{figure}[!ht]
%%% \centering
%%%	\includegraphics[width=0.32\textwidth]{img/qcd/{plooploc.D45.32.6}/{plooploc.0033.0000}.png}
%%%	\includegraphics[width=0.32\textwidth]{img/qcd/{plooploc.D45.32.6}/{plooploc.0031.0001}.png}
%%%	\includegraphics[width=0.32\textwidth]{img/qcd/{plooploc.D45.32.6}/{plooploc.0033.0015}.png}
%%% \caption{Three typical eigenmodes correlated with the local Polyakov loop at 
%%% $T=509\textmd{ MeV} \approx 2.77 T_c$. 
%%% \textit{Left}: A chiral zeromode ($\lambda=0 \textmd{ MeV}$). 
%%% \textit{Middle}: A localized nonzero mode ($\lambda = 1321.5 \textmd{ MeV} \approx 0.87 \lambda_c$). 
%%% \textit{Right}: An extended nonzero mode ($\lambda=1721.7 \textmd{ MeV}\approx 1.13 \lambda_c$).}
%%% \label{fig:ploop_localization_D45.32.6}
%%% \end{figure}

%{\bf What shall we learn about the influence of temperature ?
%Of course, the relevant non-zero eigenvalues shift upward.} 
%	Meinst Du es waere besser, nur eine Temperatur vorzufuehren?

\begin{figure}[!ht]
\centering
\includegraphics[width=\textwidth]{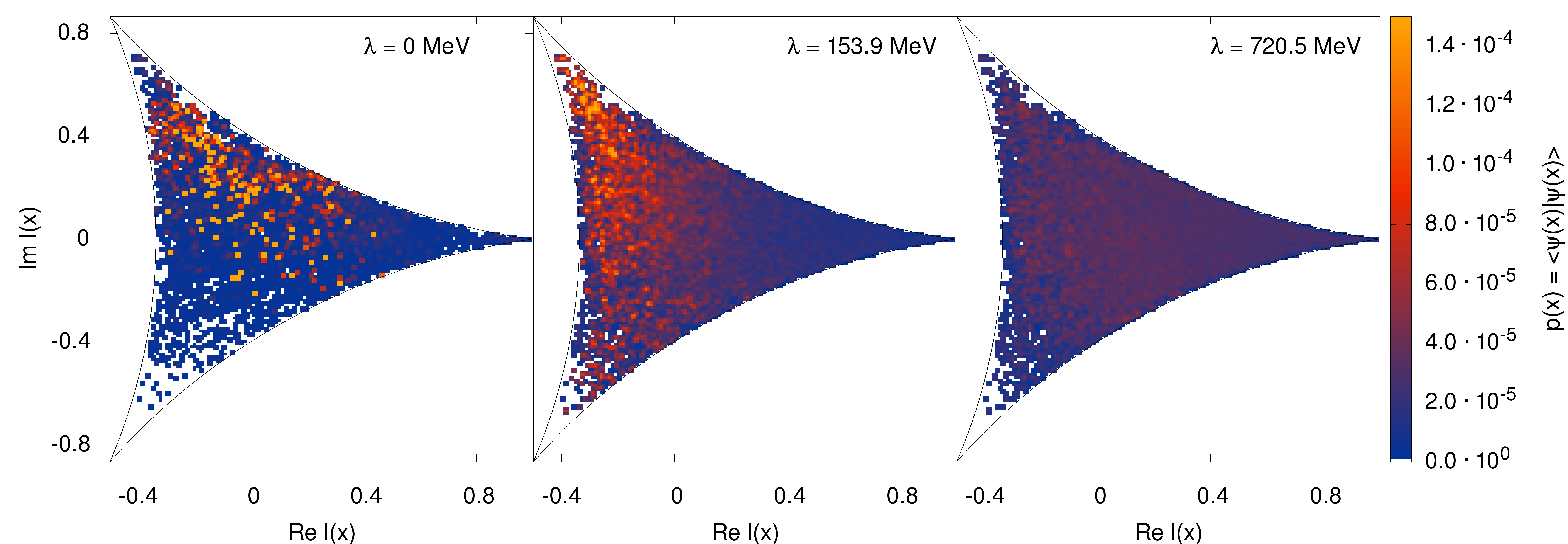}
\caption{Three typical eigenmodes located in the complex Polyakov loop plane at $T=218\textmd{ MeV} \approx 1.18 T_c$.
\textit{Left}: A chiral zeromode ($\lambda=0\textmd{ MeV}$). 
\textit{Middle}: A localized nonzero mode ($\lambda=153.9\textmd{ MeV}\approx0.85\lambda_c$). 
\textit{Right}: An extended nonzero mode ($\lambda=720.5\textmd{ MeV}\approx 3.97 \lambda_c$).}
\label{fig:ploop_localization_D45.32.14}
\end{figure}

In the complex Polyakov plane, the eigenmodes localize in regions, 
that are close to the boundary of the $l(x)$ scatter plot and the nontrivial centre 
elements of SU(3), which is shown in the left and middle panel of 
Fig. \ref{fig:ploop_localization_D45.32.14} for the temperature 
$T=218\textmd{ MeV} \approx 1.18 T_c$. Localization is reflected in a 
concentration close to regions of the plot, where two or three eigenvalues 
of the local holonomy are close to being degenerate. In the right panel of 
Fig. \ref{fig:ploop_localization_D45.32.14}, however, we see a delocalized 
mode being uncorrelated to $l(x)$.

\section{Conclusion and outlook}	
We have presented studies of the Anderson transition going on within the Dirac 
operator spectrum for overlap fermions on the background of gauge 
configurations from $N_f=2+1+1$ flavour QCD. It was found, that the temperature
dependence of the moblility edge agrees well with a linear extrapolation, and 
the localized regime disappears at the chiral crossover temperature. 
We confirmed the conjecture, that negative valued local Polyakov loop acts 
analogously to the impurities in the condensed matter Anderson model, causing 
localization.

%%% Outlook ...
In order to shed more light on the connection between mode localization and the 
topological structure of QCD, we have also studied the local gluonic topological 
charge density and its effect on Dirac eigenmode localization.
In \cite{Kovacs:2017uiz} it has been found that local 
topological objects, as described by the dilute instanton gas model, cannot 
entirely explain eigenmode localization.

\begin{figure}[!ht]
\centering
\includegraphics[width=\textwidth]{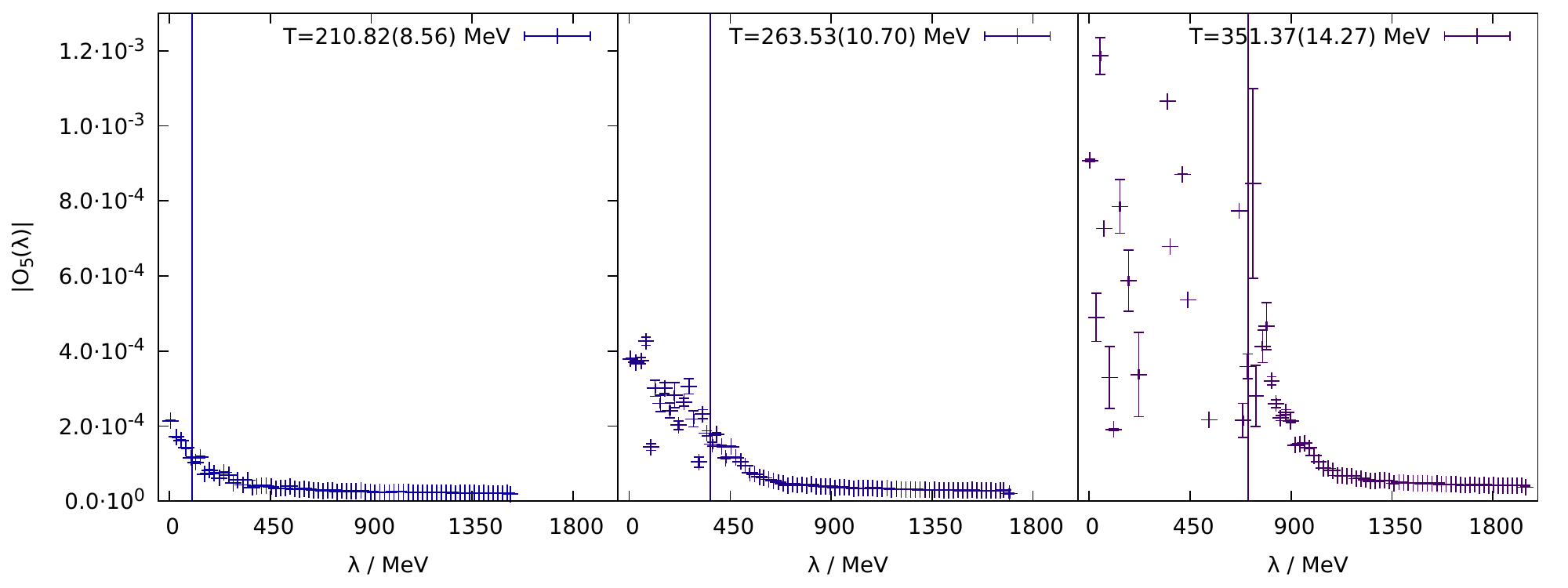}
\caption{The overlap of the gluonic topological density and the pseudoscalar density. The mobility edge is indicated with a vertical line.}
\label{fig:O5_overlap}
\end{figure}

It is, however, remarkable, that the overlap 
\begin{equation}
O_5(\lambda) = \frac{\int d^4x q(x) p_5(x)}{\frac 1 2 \int d^4x \left( (q(x))^2 + (p_5(x))^2 \right)} 
\end{equation}
between the topological density 
$q(x)=\frac{g^2}{16\pi} \epsilon_{\mu\nu\rho\sigma} F^a_{\mu\nu}(x) F^a_{\rho\sigma}(x) $ 
and the pseudoscalar density 
$p_5(x)=\langle \psi(x) | \gamma_5 | \psi(x) \rangle$ of individual modes
differs drastically between the localized and the extended spectral regions,
see Fig. \ref{fig:O5_overlap}. The difference increases with temperature.
This effect will be subject to further investigations.

%Furthermore we found, that the regions in which the low eigenmodes of the Dirac operator are localized always coincide with regions, where the (gluonic) topological charge density is large. However, we also observe regions in which $|q(x)|$ is large, but no Dirac eigenmode is localized.

%Therefore the connection between the topological charge density and the 
%localized eigenmodes remains unclear.

\bibliographystyle{unsrt}
\bibliography{references}

\end{document}